# Anion-Anion Bonding and Topology in Ternary Iridium Tin Selenides

Benjamin A. Trump,[a,b,*] Jake A. Tutmaher,[b,c] Tyrel M. McQueen[a,b,c,d]

[a]Department of Chemistry, Johns Hopkins University, Baltimore, Maryland 21218, United States
[b]Institute for Quantum Matter, Johns Hopkins University, Baltimore, Maryland 21218, United States
[c]Department of Physics and Astronomy, Johns Hopkins University, Baltimore, Maryland 21218, United States
[d]Department of Material Science, Johns Hopkins University, Baltimore, Maryland 21218, United States

*Supporting Information*

**ABSTRACT:** The synthesis and physical properties of two new and one known Ir-Sn-Se compound are reported. Their crystal structures are elucidated with transmission electron microscopy and powder X-ray diffraction. $IrSn_{0.45}Se_{1.55}$ is a pyrite phase which consists of tilted corner-sharing $IrX_6$ octahedra with randomly distributed $(Sn-Se)^{4-}$ and $(Se-Se)^{2-}$ dimers. $Ir_2Sn_3Se_3$ is a trigonally distorted skutterudite that consists of cooperatively tilted corner-sharing $IrSn_3Se_3$ octahedra with ordered $(Sn-Se)_2^{4-}$ tetramers. $Ir_2SnSe_5$ is a layered, distorted $\beta$-$MnO_2$ (pyrolusite) structure consisting of a double $IrSe_6$ octrahedral row, corner-sharing in the $a$ direction and edge-sharing in the $b$ direction. This distorted pyrolusite contains $(Se-Se)^{2-}$ dimers, $Se^{2-}$ anions, and each double row is "capped" with a $(Sn-Se)_n$ polymeric chain. Resistivity, specific heat, and magnetization measurements show that all three have insulating and diamagnetic behavior, indicative of low spin $5d^6$ $Ir^{3+}$. Electronic structure calculations on $Ir_2Sn_3Se_3$ show a *single,* spherical, non-spin-orbit split valence band, and suggest that $Ir_2Sn_3Se_3$ is topologically non-trivial under tensile strain, due to inversion of Ir-$d$ and Se-$p$ states.

## INTRODUCTION

Homologous series offer a promising opportunity for growth and design of new materials.[1] The series $M_xTCh$ is one such example, where $M$ is a late transition metal (Fe, Co, Ni, Ru, Pd, Ir, or Pt), $T$ is a later Group 14 or 15 element (Ge, Sn, Pb, As, Sb, or Bi), $Ch$ = S, Se, or Te, and $x$ = 3/2, 1, or 2/3. This series is noteworthy due to the large variety of structures that exist depending on the value of $x$ and the elements involved. These structures consist of a variety of corner- or edge-sharing $MT_3Ch_3$ octahedra (such as $FeS_2$ - pyrite) though the space groups vary wildly due to the ordering (or lack thereof) for $T$ and $Ch$. Recent studies predict a possibility of more than eight different space groups for the simple case of $x$ = 1.[2,3] When $x$ = 2/3 the skutterudite structures are commonly formed, which are of interest as promising thermoelectric materials due to their low thermal conductivity.[4,5] When $x$ = 3/2 another competing phase, half antiperovskites, are formed.[6] The structure of $M_xTCh$ depends on temperature, pressure, stoichiometry, and most notably - the transition metal itself.

Though such compounds of all of the various transition metals are structurally interesting due to anion-anion bonding, 5d transition metals have recently attracted significant interest due to strong relativistic effects (spin-orbit coupling) which could lead to non-trivial behavior.[7] These relativistic effects have comparable energy scales with crystal field stabilization and electron correlations, which could lead to magnetic frustration or possible spin liquid behavior.[8–10]

Iridium in particular has been heavily studied for these reasons, with a majority of works focusing on oxides.[11–16] Additionally, several studies were conducted on iridium chalcogenides, namely $Ir_xCh_2$ ($Ch$ = S, Se, or Te). Initial reports focused on structural details, as all of these compounds contain anion-anion bonding, and $IrS_2$, $IrSe_2$, and $IrTe_2$ can form three different structure types.[17–20] More recent investigations have been on superconductivity in both the pyrite-type $Ir_xTe_2$ ($x$ = 0.75)[21] and doped $CdI_2$-type $Ir_{1-x}M_xTe_2$ ($M$ = Pd or Pt)[22–25]. Though these studies are comprehensive, none have yet thoroughly looked at the possible stoichiometries of ternary iridium chalcogenides.

Here we report the synthesis, structure, and physical properties of the pyrite phase $IrSn_{0.45}Se_{1.55}$, the skutterudite phase $Ir_2Sn_3Se_3$, and the structurally distinct $Ir_2SnSe_5$. $Ir_2Sn_3Se_3$ has been previously reported,[4] though we expand the structural details and physical properties. To the authors' knowledge, neither $IrSn_{0.45}Se_{1.55}$ or $Ir_2SnSe_5$ have been previously reported. We find that all three exhibit insulating and diamagnetic behavior, indicative of low spin $5d^6$ $Ir^{3+}$. Each compound also displays a variation of Sn-Se bonding, as $IrSn_{0.45}Se_{1.55}$ contains Sn-Se dimers, $Ir_2Sn_3Se_3$ contains $(Sn-Se)_2$ tetramers, and $Ir_2SnSe_5$ contains $(Sn-Se)_n$ polymeric chains. Further, band structure calculations demonstrate that $Ir_2Sn_3Se_3$ is a single-band $p$- type semiconductor and imply that it becomes topologically non-trivial under tensile strain due to an inversion of Se-$p$ and Ir-$d$ states.

## EXPERIMENTAL SECTION

**Materials.** Powders were grown by placing Ir (Alfa Aesar 99.95%), Sn (Noah Technologies 99.9%), and Se (Alfa Aesar 99.999%), in stoichiometric ratios, in a fused silica tube. All tubes were filled with 1/3 atm of Ar to minimize vaporization of Sn and Se. Each tube was heated quickly to 500°C, followed by a 50°C per hour ramp to an annealing temperature at which the samples were held for four days, before being furnace cooled. The resulting boule was pulverized, pressed into a pellet, and heated at the same annealing temperature for four days, and furnace cooled again. Each resulted in a ~300 mg gray, sintered pellet which was used for all physical property and characterization methods. $Ir_2Sn_3Se_3$ was annealed at 750°C, while $Ir_2SnSe_5$ used annealed at 780°C. Later inspection indicated the presence of ~1.75 wt% $IrSe_2$ in $Ir_2SnSe_5$.

$IrSn_{0.45}Se_{1.55}$ was annealed at 950°C, and was quenched in water after each heat treatment. Targeting a 0.05 change in molar ratio resulted in significant impurities (> 10 wt%) of $IrSe_2$ or $Ir_2Sn_3Se_3$. After the second heating an Ir metal impurity around 0.15 wt% was seen which increased upon further heat treatments. The resulting pellet from $IrSn_{0.45}Se_{1.55}$ was cold-pressed rather than sintered.

**Characterization Methods.** Laboratory powder X-ray diffraction (PXRD) patterns were collected using Cu $K_\alpha$ radiation ($\lambda_{avg}$ = 1.5418 Å) on a Bruker D8 Focus diffractometer with LynxEye detector. Lebail refinements were used for phase identification and starting lattice parameters in TOPAS (Bruker AXS). Simulated annealing was then used for initial atomic positions, with Reitveld refinements for final atomic positions and lattice parameters, both in TOPAS. Synchrotron PXRD was collected on the high resolution 11-BM-B diffractometer at the Advanced Photon Source, Argonne National Laboratory, with an incident wavelength of $\lambda$ = 0.41385 Å for $Ir_2Sn_3Se_3$ and $\lambda$ = 0.41388 Å for $Ir_2SnSe_5$. Silicon was used as an internal standard for both laboratory and synchrotron PXRD; additionally 50 wt% amorphous $SiO_2$ was added to synchrotron samples to minimize absorption effects. To verify choice of lattice parameters and space group, transmission electron microscopy (TEM) was used, with a Phillips CM300 atomic resolution TEM, equipped with a Field Emission Gun with an accelerating voltage of 300 kV. For $Ir_2Sn_3Se_3$ selected area electron diffraction (SAED), collected on film (Kodak SO 163), was used to check for additional ordering. For $Ir_2SnSe_5$ SAED, collected both on film and with a CCD camera (bottom mounted Orius camera), was used to initially determine the unit cell. Structures were visualized using VESTA.[26]

Physical properties (electronic, heat capacity, thermal transport, and magnetization) data were collected on pellets in a Physical Properties Measurement System (PPMS, Quantum Design). All measurements were conducted from $T$ = 1.8 K to $T$ = 300 K. Resistivity of $Ir_2Sn_3Se_3$ was also measured down to $T$ = 70 mK on a PPMS equipped with a dilution refrigerator. All resistivity measurements used standard four-probe geometry. Heat capacities were measured using the semi-adiabatic pulse technique, with three repetitions at each temperature. Magnetic susceptibilities were measured with a $\mu_0H$ = 1 T.

**Calculation Methods.** Electronic and band structure calculations were performed on $Ir_2Sn_3Se_3$, using density functional theory (DFT) with the local density approximation (LDA) utilizing the ELK all electron full-potential linearized augmented-plane wave plus local orbitals (FP-LAPW+LO) code.[27] Calculations were conducted both with and without spin-orbit coupling (SOC) using a 4 x 4 x 4 k-mesh, with the experimental unit cell. Parity analysis on the time-reversal invariant momentum (TRIM) points for $Z_2$ values[28] were conducted by fitting the eigenvectors computed from ELK to a set of maximally-localized Wannier functions (MLWF, using Wannier90 software package[29]). The calculation on $Ir_2Sn_3Se_3$ under tensile strain was conducted using spin-orbit coupling and a unit cell increased uniformly by 0.6 Å. The band structures were independently verified using the Vienna Ab Initio Simulation Package (VASP)[30–32], and the topological indices were alternatively calculated using MLWFs generated by Wannier90 in tandem with the Z2Pack software[33,34].

## RESULTS AND DISCUSSION

**Structure of $IrSn_{0.45}Se_{1.55}$.** Room temperature laboratory PXRD data for $IrSn_{0.45}Se_{1.55}$ is shown in Figure 1a. Refinements were conducted with space group $Pa\bar{3}$, the model structure is shown in the inset of Figure 1a. Crystallographic parameters are in Table S1. The structure type is identical to the pyrite $FeS_2$[35], with tilted, corner-sharing Ir$X_6$ octahedra, and Se-Se dimers on each corner, with Sn randomly distributed over the Se sites. Similar compounds such as cobaltite $CoAsS$[36] or ullmannite $NiSbS$[37] show anion ordering leading to a lower symmetries of $Pca2_1$ and $P2_13$ respectively. Ordering in these compounds is justified due to the observation of the (010) reflection for ullmannite and additionally the (110) reflection for cobaltite.[36] However, PXRD for $IrSn_{0.45}Se_{1.55}$ does not show the evidence of either of these reflections despite as much as 55,000 counts for peaks, and a strip detector with a high signal to noise ratio. This defends the choice of space group $Pa\bar{3}$. Furthermore, attempts to refine occupancies led to values within 1% of nominal stoichiometry for $IrSn_{0.45}Se_{1.55}$.

This stoichiometry is very close to $Ir_2SnSe_3$. Given the samples are diamagnetic (see Figure S1), and thus Ir is in the 3+ oxidation state, this implies a mixture of Sn-Se and Se-Se, i.e. $Ir^{3+}_2(SnSe)^{4-}(Se_2)^{2-}$, but with a slight excess of Se-Se dimers. This deviation from "perfect" stoichiometry is an explanation for the lack of ordering in $IrSn_{0.45}Se_{1.55}$. Attempts to target $IrSn_{0.5}Se_{1.5}$ were unsuccessful, resulting in a ~21 wt% Ir$_2$Sn$_3$Se$_3$ impurity, while attempts to target IrSn$_{0.4}$Se$_{1.6}$ had a ~12 wt% IrSe$_2$. In other words, accessing stoichiometric Ir$_2$SnSe$_3$ was not possible under our conditions, implying that it is less thermodynamically stable than competing phases. However, the off stoichiometric IrSn$_{0.45}$Se$_{1.55}$ is accessible as the tail of a Ir$_2$Sn$_{1-\delta}$Se$_{3+\delta}$ solid solution because it lies outside the phase field of the completing more stable products.(see Figure S2).

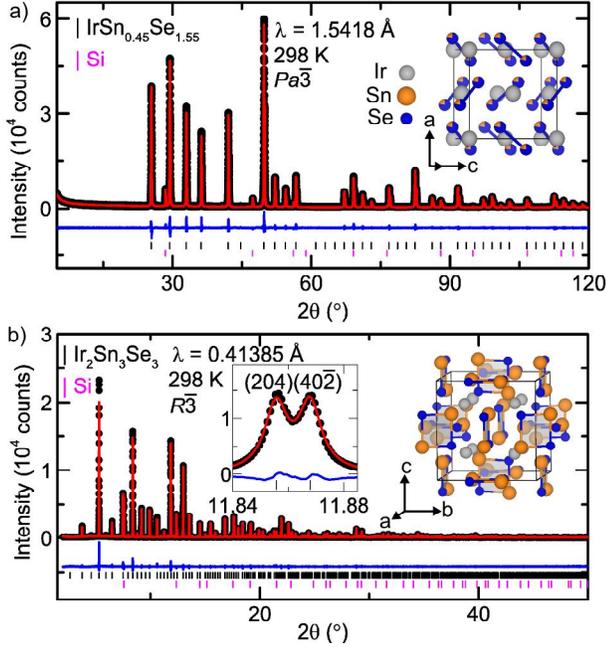

**Figure 1.** a) Reitveld refinement of laboratory powder X-ray data for IrSn$_{0.45}$Se$_{1.55}$ with internal Si standard. Light blue arrow shows a peak for 0.15 wt% Ir metal impurity. Structure is shown in the inset, which models Sn randomly mixed on the Se sites. b) Reitveld refinement of synchrotron powder X-ray data for Ir$_2$Sn$_3$Se$_3$ with internal Si standard. Insets show (left) subtle splitting of peaks and (right) Ir$_2$Sn$_3$Se$_3$ shown as Sn$_2$Se$_2$ tetramers. Experimental data shown as black circles, fit is in red, with the difference in blue. Ir is shown in grey, Sn in orange, and Se in blue.

**Structure of Ir$_2$Sn$_3$Se$_3$.** Figure 1b shows synchrotron PXRD data for Ir$_2$Sn$_3$Se$_3$ using space group $R\bar{3}$. It was previously reported as a skutterudite. The prototypical $Im\bar{3}$ skutterudite is CoAs$_3$ which forms square (As$_4$)$^{4-}$ tetramers.[4] There is a clear splitting of the (204) and (402) reflections shown in the left inset of Figure 1b. It is well known that changing the formula of skutterudites to M$_{2/3}$TCh can lead to anion ordering, resulting in distorted (T-Ch)$_2$ tetramers and a reduction in crystallographic symmetry to $R\bar{3}$. Using this as a starting model, we are able to obtain an excellent fit of the model to the data. The experimental trigonal unit cell is also within 0.13% difference with the previously reported cubic structure for IrSn$_{1.5}$Se$_{1.5}$.[4] Furthermore, SAED (not shown) does not indicate any doubling of the unit cell, or any other ordering, hence the unit cell and space group are well justified. Crystallographic parameters are in Table S2. Attempts to refine occupancies led to values within 1% of unity. The right inset in Figure 1b demonstrates the structure consists of (Sn-Se)$_2$ tetramers or distorted squares. The electron count can be understood as Ir$^{3+}_4$(Sn$_2$Se$_2$)$^{4-}_3$, which chemically compares well to other skutterudites.[4]

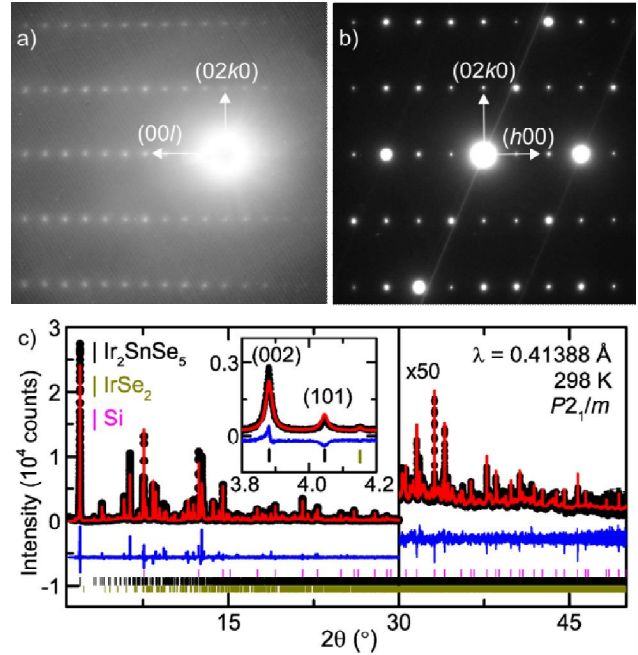

**Figure 2.** Selected area electron diffraction for Ir$_2$SnSe$_5$ along a) ($h$00) and b) (00$l$) planes. c) Rietveld refinement of synchrotron powder X-ray data with internal Si standard. Experimental data shown as black circles, fit is in red, with the difference in blue. Inset shows that the model over-fits a diagonal (101) plane and under-fits the (002) plane. Contribution of ~1.75 wt% IrSe$_2$ impurity is also seen.

**Structure of Ir$_2$SnSe$_5$.** SAED patterns oriented in ($h$00) and (00$l$) directions are shown in Figure 2a and Figure 2b respectively for Ir$_2$SnSe$_5$. In the $y$ direction the spacing in both patterns is directly related to the $b$ lattice parameter, while the $x$ direction is directly related to the $c$ and $a$ lattice parameters for the ($h$00) and (00$l$) patterns respectively. Figure 2c shows the room temperature synchrotron PXRD data for Ir$_2$SnSe$_5$ using space group $P2_1/m$. The corresponding Rietveld refinement included a 1.75 wt% IrSe$_2$ impurity, as well as an internal Si standard. Crystallographic parameters are shown in Table S3.

The choice of space group and unit cell are justified through the SAED patterns and the experimental synchrotron PXRD data. Hamilton R-ratio tests[38] and $\chi^2$ ratio tests[39] against other space groups ($P1$, $P\bar{1}$, $P2$, $P2_1$, $Pm$, $P2/m$, $P2_1/m$:2) confirms, with 99% confidence, the choice of space group $P2_1/m$. Additionally, tests using ADDSYM in PLATON[40] did not find any additional symmetry. The lattice parameters determined from SAED are within 10% difference of those reported in Table S3, from Rietveld refinement, likewise the SAED patterns also do not show any evidence of additional order or doubling of the unit cell. Lastly, all observed peaks are fit by this model, and our model distinguishes between Sn and Se as Hamilton R-ratio tests[38] and $\chi^2$ ratio tests[39] for alternative Sn posi-tions shows 99.99% confidence of our proposed Sn position.

Nonetheless, the fit in Figure 2c is visibly imperfect due to lower angle peaks that are severely under-fit. This is highlighted by the inset in Figure 2c where the model is seen to under-fit for the (002) reflection and over-fit for the (101) reflection. Systematically the model over-fits diagonal reflections (e.g. (110), (103), etc.) and under-fits other reflections (e.g. (002), (020), (100), etc.). These systematic deviations, along with the certainty in the unit cell, space group, and atomic positions then suggests a stacking fault in the $c$ direction. Careful observation of the $Ir_2SnSe_5$ structure in Figure 3 demonstrates that a shift in the $b$ direction could exist due to the Van der Waals gap. This shift would cause diagonal reflections across layers to broaden and have less intensity, while maintaining the sharpness and intensity of reflections that exist within each layer (e.g. (100), (020), (002), etc.). Thus we propose the imperfections of our model in describing the data are due to a stacking fault in the $c$ direction.

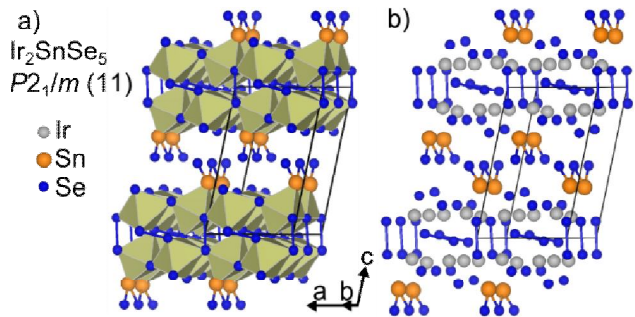

**Figure 3.** Structure of $Ir_2SnSe_5$ just off the $ac$ plane a) highlighting corner-sharing in the $ac$ plane, edge-sharing in the $bc$ plane, and b) both Se-Se dimers and the $(Sn-Se)_n$ polymeric chain. Ir is shown in gray, Sn in orange, and Se in blue.

The proposed structure, shown in Figure 3, of the layered, distorted $\beta$-$MnO_2$ (pyrolusite)[41] type. Each layer contains a double $IrSe_6$ octahedral row, corner-sharing in the $ac$ plane and each row is edge-sharing in the $bc$ plane. This structure bears similarity to the $IrSe_2$ structure, a three-dimensional structure that contains both a pyrolusite and ramsdellite portions.[18] Both $Ir_2SnSe_5$ and the pyrolusite portion of $IrSe_2$ contain $(Se_2)^{2-}$ anion dimers stabilizing the octahedra. $Ir_2SnSe_5$ is structurally distinct however, as it also contains a $(Sn-Se)_n$ polymeric chain "capping" each double octahedral layer. In other words, $IrSe2$ can be structurally described as $Ir^{3+}_2(Se_2)^{2-}Se^{2-}_2$,[18] $Ir_2SnSe_5$ is the same, but with the addition of a charge-neutral Sn-Se polymeric chain, i.e. $Ir^{3+}_2(Se_2)^{2-}Se^{2-}_2(SnSe)^0$.

**Sn-Se bonding in the Ir-Sn-Se system.** Each Ir-Sn-Se compound contains some form of anion-anion, or Zintl-like bonding. This is common for Ir compounds, as $IrCh_2$ ($Ch$ = S, Se, or Te) compounds all have similar effects.[18–20] Figure 4 highlights the difference in Sn-Se anion-anion bonding in each Ir-Sn-Se compound. $IrSn_{0.45}Se_{1.55}$ contains both $(Se-Se)^{2-}$ and $(Sn-Se)^{4-}$ dimers, with an average distance of 2.652(1) Å. $Ir_2Sn_3Se_3$ contains $(Sn-Se)_2$ tetramers instead, with a long and short distance of 2.868(5) Å and 2.68(1) Å. This is exactly what is expected if two $(Sn-Se)^{4-}$ dimers are joined together along with the removal of four electrons.

The bond distances in the Sn-Se polymeric chain are shorter than the monomer and dimer, which indicates more ionic character. This is expected from an electron counting argument, as the valence for Sn is formally -2, 0, and +2 for $(Sn-Se)^{4-}$ dimers, $(Sn-Se)_2^{4-}$ tetramers, and $(Sn-Se)_n$ respectively, following a trend of Zintl-like bonding to more ionic type bonding.

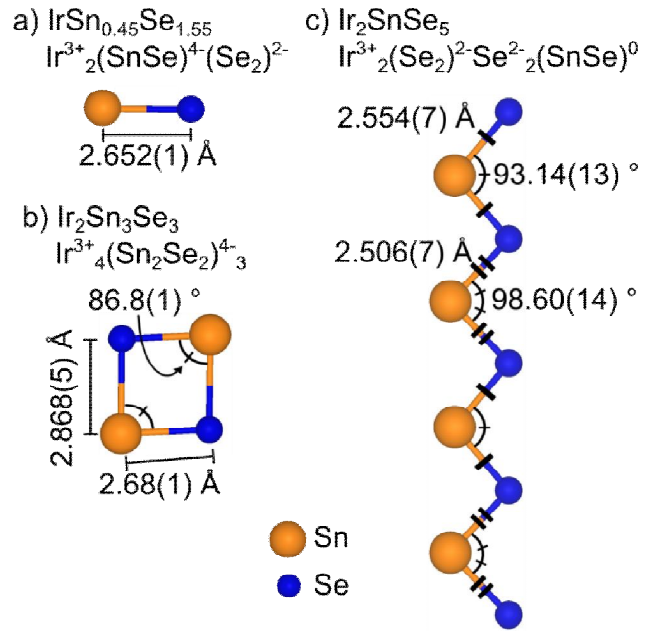

**Figure 4.** a) The Sn-Se dimer in $IrSn_{0.45}Se_{1.55}$. The distance given is an average for $Se_2$ and Sn Se dimers. b) The $(Sn-Se)_2$ tetramer in $Ir_2Sn_3Se_3$. c) The $(Sn-Se)_n$ polymeric chain in $Ir_2SnSe_5$.

**Physical Properties.** Figure 5a shows the heat capacity for $Ir_2SnSe_5$, $IrSn_{0.45}Se_{1.55}$, and $Ir_2Sn_3Se_3$ as $C_p/T^3$ vs log$T$ to highlight acoustic and optic phonon modes.[42] Plotted in this way Einstein (optic) modes, appear as a peak, while Debye (acoustic) modes, increase upon cooling until becoming constant. Additionally, electronic heat capacity appears as a sharp increase at low temperatures. Scaling all three data sets by the amount of amounts indicates that all three have a similar Debye and a similar Einstein mode. This Einstein mode appears to decrease in energy from $IrSn_{0.45}Se_{1.55}$ to $Ir_2Sn_3Se_3$ (red and blue arrows respectively), and contributes even less to $Ir_2SnSe_5$. Meanwhile the contribution of a Debye mode appears to increase from $IrSn_{0.45}Se_{1.55}$, to $Ir_2Sn_3Se_3$, to $Ir_2SnSe_5$. Both observations are consistent with the change in Sn-Se bonding, which goes from dimers, to tetramers, to polymeric chains. As the connectivity of the dimers increases, their dimensionality increases, and their associated modes broaden and appear more Debye-like, as seen for the heat capacity of $Ir_2SnSe_5$ which is almost purely Debye-like. The small feature at $T \sim 4.5$ K for all three is due to helium condensation around this temperature.

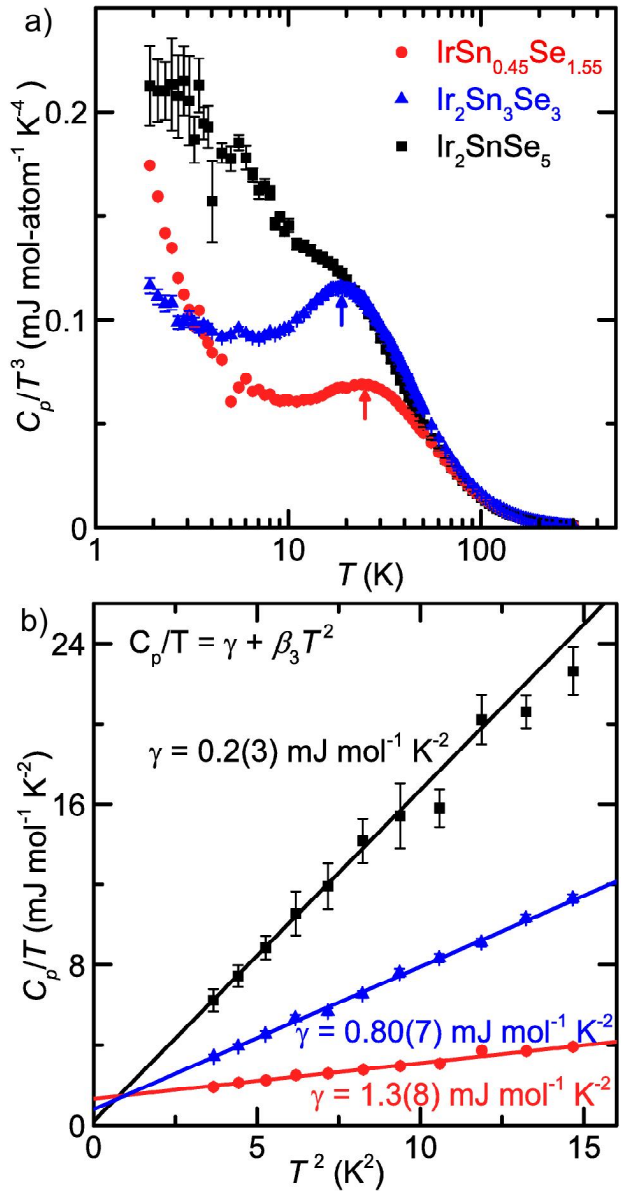

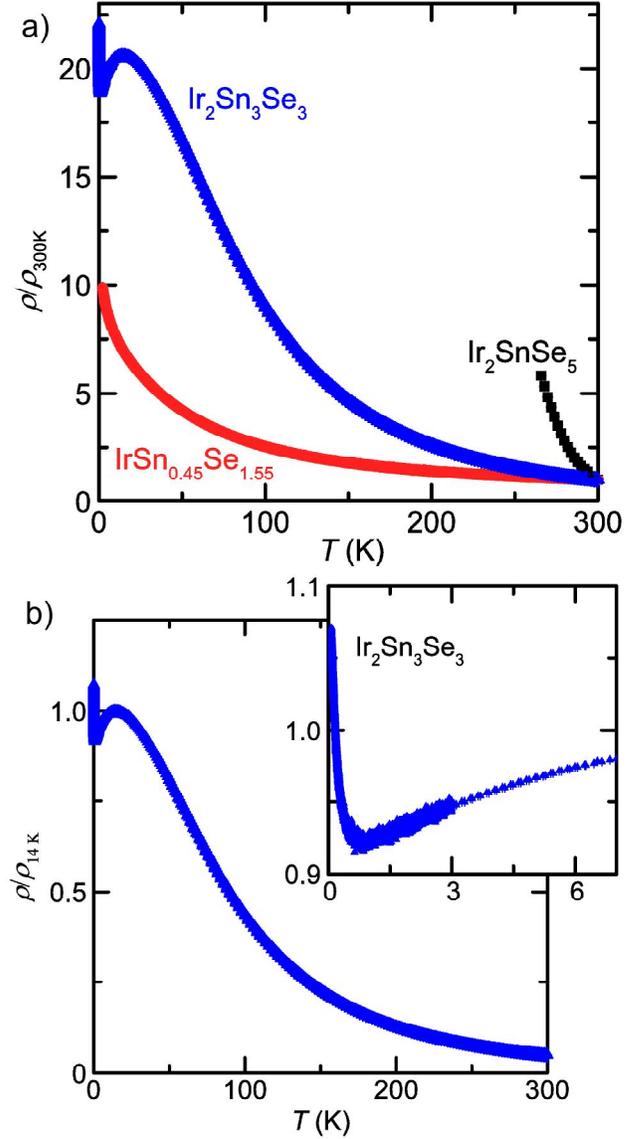

**Figure 5.** a) Heat capacity over temperature cubed versus log of temperature for $Ir_2SnSe_5$ (black squares), $IrSn_{0.45}Se_{1.55}$ (red circles), and $Ir_2Sn_3Se_3$ (blue triangles), scaled per atom, emphasizing an Einstein mode that shifts to lower energy (red and blue arrows). b) Heat capacity over temperature versus temperature squared highlighting the electronic heat capacity ($\gamma$). Solid lines are fits extrapolated to zero.

The plot of $C_p/T$ vs $T^2$ in Figure 5b shows the relationship between the electronic ($\gamma$) and phonon ($\beta_3$) contributions to specific heat, fit to the equation $C_p/T = \gamma + \beta_3 T^2$.[43] $Ir_2SnSe_5$ has an electronic specific heat which within error of zero (0.2(3) mJ K$^{-2}$ mol$^{-1}$), while $Ir_2Sn_3Se_3$ is has a non-zero $\gamma$ of 0.80(7) mJ K$^{-2}$ mol$^{-1}$. The small, non-zero $\gamma$ for $Ir_2Sn_3Se_3$ is in agreement with the low temperature ($T < 2$ K) upturn in $C_p/T^3$ in Figure 5a. Similarly, $IrSn_{0.45}Se_{1.55}$ also has a non-zero $\gamma = 1.3(8)$ mJ K$^{-2}$ mol$^{-1}$ which is also in agreement with the sharp increase in low temperature ($T < 5$ K) $C_p/T^3$ in Figure 5a.

**Figure 6.** a) Normalized resistivity as a function of temperature for $Ir_2SnSe_5$ (black squares), $IrSn_{0.45}Se_{1.55}$ (red circles), and $Ir_2Sn_3Se_3$ (blue triangles). Errors are contained in the size of the symbols. b) Normalized for $Ir_2Sn_3Se_3$ shows a broad feature at $T = 40$ K and then an increase again at $T = 0.75$ K.

Normalized resistivity for $Ir_2SnSe_5$, $IrSn_{0.45}Se_{1.55}$, and $Ir_2Sn_3Se_3$ is shown in Figure 6. The rate at which the normalized resistivity increases at lower temperatures is proportional to the size of its' semiconducting gap, which increases from $IrSn_{0.45}Se_{1.55}$, to $Ir_2Sn_3Se_3$, to $Ir_2SnSe_5$. $Ir_2SnSe_5$ could not be measured below $T = 266$ K due to the large resistivity, which agrees with the zero electronic contribution of specific heat in Figure 7b. Though the insulating resistivity of $IrSn_{0.45}Se_{1.55}$ is conflicting with the heat capacity in Figure 7, it is not uncommon for a semiconductor to have a non-zero $\gamma$ due to a finite doping (carrier concentration). $Ir_2Sn_3Se_3$ is also semiconducting with a non-zero $\gamma$, however it is accompanied by unique low temperature ($T > 40$ K) behavior.

The normalized resistivity for $Ir_2Sn_3Se_3$, shown in Figure 6b, follows insulating behavior until $T \sim 40$ K, where it first appears to plateau, then decreases, until finally increasing again at $T \sim 0.75$ K. Though there are many complex explanations for this phenomenon, this type of behavior is well known in heavily doped semiconductors, and has been observed in p-type Ge [44,45] and modeled in p-type Si [46] for similar carrier concentrations.

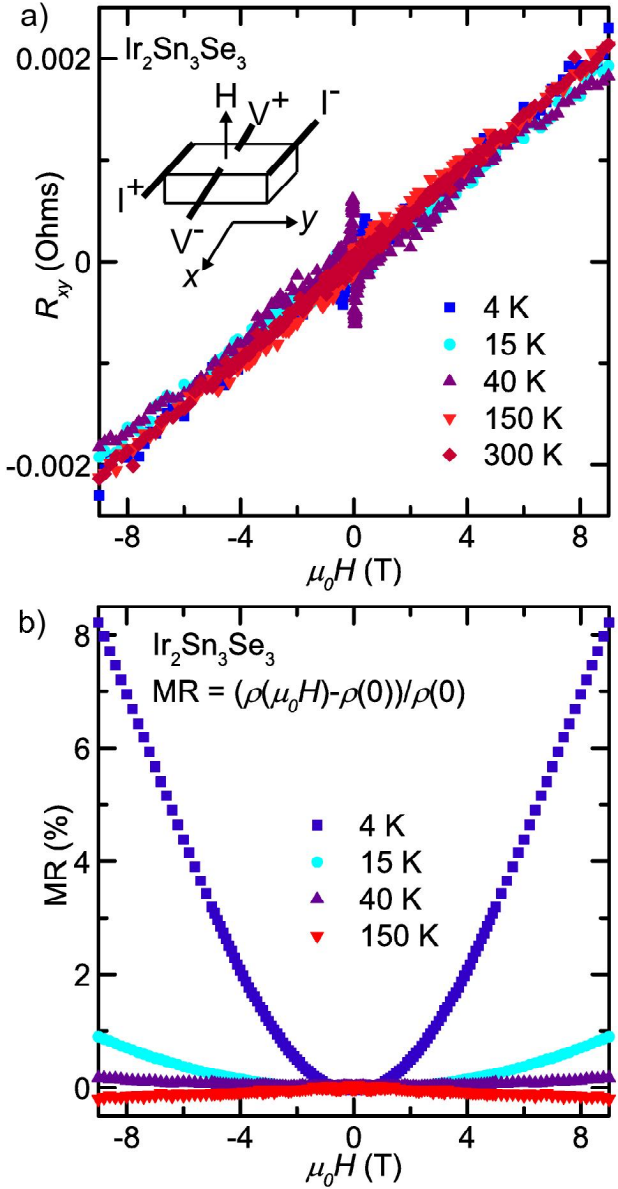

**Figure 7.** a) Hall resistance versus applied field for $Ir_2Sn_3Se_3$ at various temperatures. The inset shows the experimental setup. b) Magnetoresistance (MR) of $Ir_2Sn_2Se_3$ as a function of applied field. The magnitude changes sign around $T = 40$ K and increases as temperature decreases. Errors are contained by the size of the symbols for both.

Hall resistance ($R_{xy}$) measurements as a function of applied field are shown in Figure 7a for several temperatures ($T = 4, 15, 40, 150, 300$ K). The sample setup is shown in the inset in Figure 7 a and the data was symmeterized. The slope of these lines, normalized by sample thickness ($d$), give the hall coefficient ($R_H$), which is equal to $1/ne$ where $n$ is the carrier concentration and $e$ is the elementary charge.[47] All temperatures have a similarly positive slope, indicative of p-type doping and a roughly temperature-independent carrier concentration of $2.2(2)*10^{19}$ cm$^{-3}$. This carrier concentration agrees with the previously reported $2.3*10^{19}$ cm$^{-3}$,[4] the resistivity in Figure 6, and the small, non-zero $\gamma$ in Figure 5. Both p-type Ge and Si show similar resistivity versus temperature behavior for a carrier concentration $\sim 10^{19}$ cm$^{-3}$, and an electronic specific heat is expected. Semiconducting behavior is also expected as the carrier concentration is still below the Mott metal to insulator transition ($n < \sim 10^{22}$ cm$^{-3}$).[48,49]

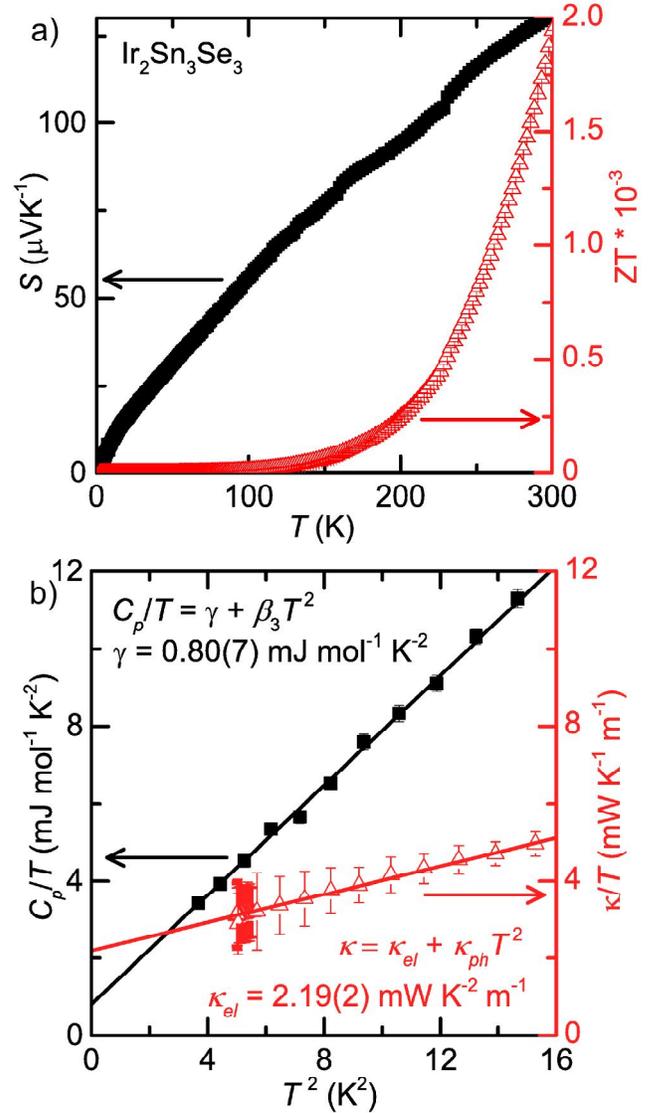

**Figure 8.** a) Seebeck coefficient (black squares) and the dimensionless ZT figure of merit (red triangles) as a function of temperature for $Ir_2Sn_3Se_3$. b) Thermal conductivity (red triangles) and heat capacity (black squares), both over temperature, as a function of $T^2$ to separate lattice and electronic contributions.

There is a small apparent increase in the carrier concentration from Hall measurements from $2.01(2)*10^{19}$ cm$^{-3}$ to $2.40(2)*10^{19}$ cm$^{-3}$, from $T = 300$ K to $T = 15$ K respectively. This is most logically explained as arising from ther-mally excited states across the gap at high temperature. This explanation is consistent with the normalized resistivity for $Ir_2Sn_3Se_3$, which increases until becoming constant at $T = 15$ K, where the intrinsic gap no longer contributes thermal n-type carriers. At $T < 15$ K the carrier concentration decreases to $1.91(2)*10^{19}$ cm$^{-3}$ at $T = 4$ K as extrinsic p-type carriers begin to decrease. Attempts to fit this data with a simple multi-gap model, as done by Fritzsche,[44] have proved unsuccessful. A successful model to this data would need to include not only impurity scattering (neutral and charged), lattice scattering, and hole-hole scattering for both intrinsic and extrinsic carriers as done by Li [46]; but this model would also need to include grain effects due to a sintered sample. The complexity of a model that could successfully describe these effects is beyond the scope of this manuscript.

Hall mobilities were also calculated using $\mu_H = R_H/\rho_{xx}$, with the values shown in Table S4. These values are roughly in agreement with those calculated from resistivity alone ($\rho^{-1} = en\mu$), also shown in Table S4. The above relationship between resistivity and mobility also explains why the mobilities decrease as temperature decreases, as the carrier concentration is roughly temperature-independent. The mobility values are also reasonable considering the semiconducting behavior of $Ir_2Sn_3Se_3$, and the Hall mobilties are within 3% difference from those previously reported in Ref. 4. Both resistivity and Hall measurements probe the experimental mobility, with values that are an "average" of extrinsic and intrinsic carriers. It is likely that a probe which better measures only the extrinsic carriers would yield a larger mobility.

Magnetoresistance (MR) for $Ir_2Sn_3Se_3$ is shown in Figure 7b for various temperatures. MR is commonly defined as $(\rho(H)-\rho(0))/\rho(0)$ and given as a percent. This data was collected using the sample setup shown in the inset of Figure 7a and was symmeterized. For single-carrier semiconductors MR is positive and follows a $1+(\mu H)^2$ trend, where $\mu$ is the mobility and $H$ is the applied field.[50] $Ir_2Sn_3Se_3$ appears to not follow these trends for several reasons. The MR for $Ir_2Sn_3Se_3$ is only positive when $T \leq 40$ K, which is where the resistivity begins to plateau (Figure 6b), and the intrinsic gap no longer contributes carriers. Secondly, MR for $T = 4$ K appears to follow Kohler's rule (MR $= 1+(\mu_{MR}H)^2$)[50] initially, with $\mu_{MR} = 366(3)$ cm$^2$V$^{-1}$s$^{-1}$, but then appears to have a linear relationship above $\mu_0 H = 3$ T. This deviation could arise for a myriad of reasons, such as the polycrystalline nature of the sample, different carrier types, or a function of anisotropy.[50] The magnitude of $\mu_{MR}$ is also much higher than that from resistivity and Hall measurements, however this is likely due to this mobility being from extrinsic carriers alone, which would have a higher mobility than intrinsic carriers. Single crystal studies are necessary to truly determine the origin of the linear MR behavior at higher fields.

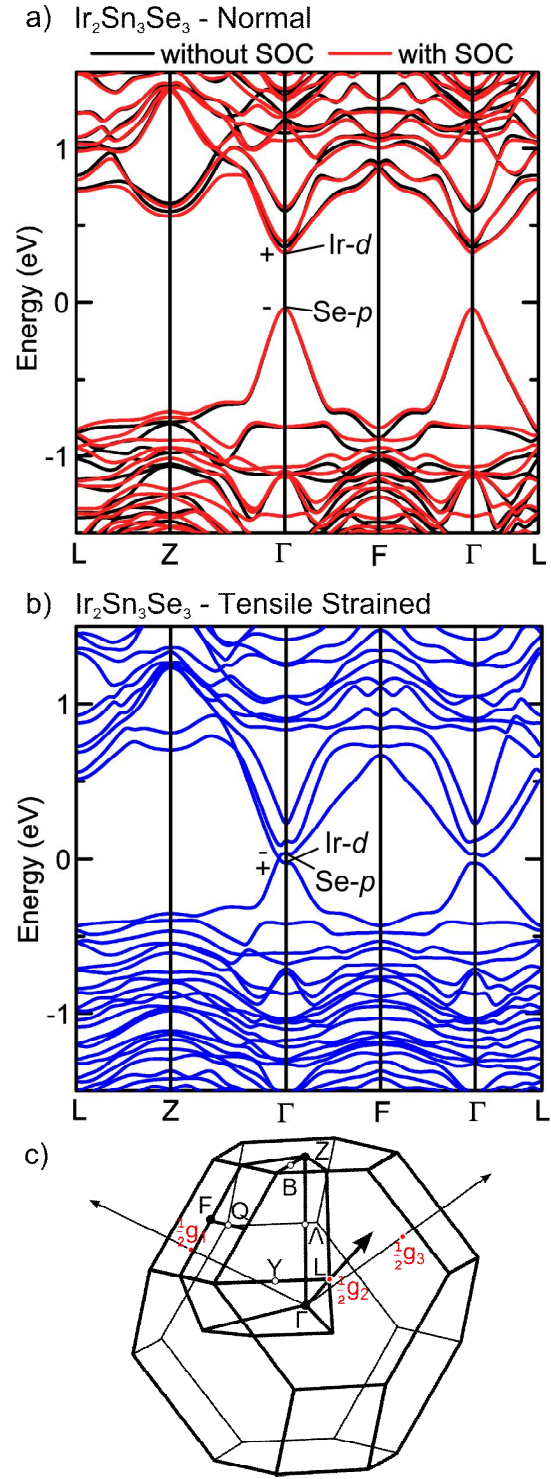

**Figure 9.** a) Band structure for $Ir_2Sn_3Se_3$ without (black) and with (red) spin-orbit coupling (SOC). Ir-d states are seen just above the Fermi level, while Se-p states are just below. b) Tensile strained band structure for $Ir_2Sn_3Se_3$, using a unit cell uniformly expanded unit cell by 0.6 Å. The Ir-p and Se-d states at the Γ point invert. c) The Brillioun zone for R3 $Ir_2Sn_3Se_3$ with special points and reciprocal lattice vectors shown.

Results from thermal transport measurements are shown in Figure 8a for the skutterudite $Ir_2Sn_3Se_3$. The left axis of Figure 8a shows the Seebeck coefficient (S) versus temperature, and appears positive and roughly linear. Noting that $S \propto m^* n^{-2/3} T$,[51] this suggests that the carrier concentration ($n$) and enhanced mass ($m^*$) are roughly temperature independent. The sign of the $S$ indicates $Ir_2Sn_3Se_3$ is hole doped, in agreement with the Hall measurements in Figure 7, which also indicates $p$-type doping and a roughly temperature-independent carrier concentration. The right axis of Figure 8a shows the dimensionless ZT figure of merit, with ZT = $S^2\sigma T/\kappa$.[51] The shape of the curve is well understood, as it follows the trends of $S^2$ and the inverse of the resistivity for $Ir_2Sn_3Se_3$.

**Calculations on $Ir_2Sn_3Se_3$.** Calculations on $Ir_2Sn_3Se_3$ were run to further understand the electronic and transport properties, shown in Figure 9. The band structure in Figure 9a, with the associated Brillouin zone in Figure 9c, implies that despite the complex structure, $Ir_2Sn_3Se_3$ is a semiconductor, with a direct gap of ~0.4 eV at the Γ point and a single valence band. Ir-$d$ and Se-$p$ states contribute to the bands directly above and below the Fermi level respectively. Such a simple band structure is unexpected from the structure, which warrants further investigation of the bands at the Γ point.

High pressure (compressive strain/reduced unit cell) calculations were conducted, which led to an unexpected *increase* in the band gap. This led to reduced pressure calculations (tensile strain), using a uniformly 0.6 Å expanded unit cell (a 6.7% increase), with the band structure shown in Figure 9b. This reduction of the unit cell led to an inversion of the Ir-$d$ and Se-$p$ states at the Γ point. The $Z_2$ topological invariant was calculated at each of the TRIM points, which for $R\overline{3}$ are Γ, 3F, 3L, and Z. Multiplying the parity from the occupied states at each TRIM, then multiplying the parity of the TRIM points, reveals that $Ir_2Sn_3Se_3$ is a topologically trivial ($Z_2$ = +1), while tensile strained $Ir_2Sn_3Se_3$ is topologically nontrivial ($Z_2$ = -1) due to the inversion of Ir-$d$ and Se-$p$ states at the Γ point. Additional calculations show that Ir2Sn3Se3 has a topological index of 1:(0,0,0), consistent with a strong topological insulator with a band inversion at Γ.

Using the determined carrier concentration of $n$ = 2.2(2)*$10^{19}$ cm$^{-3}$, the electronic specific heat of $\gamma$ = 0.80(7) mJ mol$^{-1}$ K$^{-2}$, along with the electronic band structure and density of states (DoS) calculations, we determine an enhanced mass ($m^*/m$) of 7.0(4). This then allows us to calculate a mobility from $\mu = e\tau/m^*$, and the relationship between heat capacity ($C$) and thermal conductivity ($\kappa$), $\kappa = \frac{1}{3}Cv^2\tau$.[47] The respective electronic heat capacity and thermal conductivity is shown in Figure 8b, plotted versus $T^2$ to separate the electronic ($\gamma$ and $\kappa_{el}$) and phonon ($\beta_3$ and $\kappa_{ph}$) contributions.[43,52] This leads to a mobility of 1110(60) cm$^2$V$^{-1}$s$^{-1}$, a value much larger than the experimental mobilities calculated from resistivity and Hall measurements. This difference is due to what each experimental probe measures - resistivity and Hall measurements probe an "average" for all types of carriers, both intrinsic and extrinsic. Electronic heat capacity and electronic thermal conductivity are only from the number of carriers at the Fermi level, which are extrinsic $p$-type carriers due to defects. These extrinsic carriers would have a much larger mobility than thermal intrinsic carriers across a ~0.4 eV gap. Therefore we propose the mobilities of 366(3)-1110(60) cm$^2$V$^{-1}$s$^{-1}$ are from extrinsic carriers and the mobilities of 0.1-10 cm$^2$V$^{-1}$s$^{-1}$ are from both extrinsic and intrinsic carriers.

## CONCLUSION

The structures of $IrSn_{0.45}Se_{1.55}$, $Ir_2Sn_3Se_3$, and $Ir_2SnSe_5$ are reported. $IrSn_{0.45}Se_{1.55}$ is a new $Pa\overline{3}$ pyrite phase with randomly distributed Sn-Se and Se-Se dimers. $Ir_2Sn_3Se_3$ is a trigonally-distorted $R\overline{3}$ skutterudite, and the lattice parameter is within 0.13% of that previously reported cubic structure[4]. $Ir_2SnSe_5$ is a new, layered, $\beta$-MnO$_2$-like structure, containing double octahedral $IrSe_6$ rows, corning-sharing in the $a$ direction, and edge-sharing in the $b$ direction, with each double octahedral layer effectively "capped" by (Sn-Se)$_n$ polymeric chains.

Electron counting suggests formulas of $Ir^{3+}{}_2(SnSe)^{4-}(Se_2)^{2-}$ for $IrSn_{0.45}Se_{1.55}$, $Ir^{3+}{}_4(Sn_2Se_2)^{4-}{}_3$ for $Ir_2Sn_3Se_3$, and $Ir^{3+}{}_2(Se_2)^{2-}Se^{2-}{}_2(SnSe)^0$ for $Ir_2SnSe_5$ with Sn-Se dimers, (Sn-Se)$_2$ tetramers, and (Sn-Se)$_n$ polymeric chains respectively. The anion anion bonding is consistent with other Ir chalcogenides.[18–20] All three compounds are insulating and diamagnetic indicative of $5d^6$ $Ir^{3+}$.

Hall measurements, thermal transport, heat capacity, and resistivity measurements, as well as electronic structure calculations, on $Ir_2Sn_3Se_3$ demonstrate $p$-type doping with a carrier concentration of 2.2(2)*$10^{19}$ cm$^{-3}$, an enhanced mass ($m^*/m$) of 7.0(4), electronic specific heat ($\gamma$) of 0.80(7) mJ mol$^{-1}$ K$^{-2}$, electronic thermal conductivity ($\kappa_{el}$) of 2.19(2) mW K$^{-2}$ m$^{-1}$. Heat capacity measurements also show an Einstein mode which broadens and becomes more Debye-like as the dimensionality of anion-anion bonding increases. Mobilities of 366(3)-1120(6) cm$^2$V$^{-1}$s$^{-1}$ were determined for extrinsic $p$-type carriers, using magnetoresistance, heat capacity, and thermal conductivity. Resistivity and Hall measurements also measured mobilities of 0.1-10 cm$^2$V$^{-1}$s$^{-1}$ which represent the overall mobility of both extrinsic and intrinsic carriers.

Electronic band structure calculations also reveal that $Ir_2Sn_3Se_3$ is a direct-gap semiconductor with a gap of ~0.4 eV. Uniformly expanding the unit cell by 0.6 Å appears to turn $Ir_2Sn_3Se_3$ into a topological insulator. Though this 6.7% enlargement is moderately sizable, such an increase may be possible with strain and/or similar compounds with larger atoms.

## ASSOCIATED CONTENT

**Supporting Information.**
The Supporting Information is available free of charge on ACS Publications website at DOI:   .
    Magnetization versus temperature for $Ir_2Sn_{0.45}Se_{1.55}$, $Ir_2Sn_3Se_3$, and $Ir_2SnSe_5$, ternary Ir-Sn-Se diagram, crystallographic information for $Ir_2Sn_{0.45}Se_{1.55}$, $Ir_2Sn_3Se_3$, and $Ir_2SnSe_5$, and mobility values for $Ir_2Sn_3Se_3$. (PDF)

**AUTHOR INFORMATION**

**Corresponding Author**
*E-mail: btrump1@jhu.edu.

**Notes**
The authors declare no competing financial interests.

**ACKNOWLEDGMENT**

This work was supported by NSF, Division of Materials Research (DMR), Solid State Chemistry (SSMC), CAREER grant under Award DMR-1253562, and the ICDD Ludo Frevel Crystallography Scholarship. Use of Dilution Refrigerator funded by National Science Foundation Major Research Instrumentation Program under NSF DMR-0821005. B.A.T. also acknowledges useful discussions with A.M. Fry, W.A. Phelan, and K. JT Livi.

REFERENCES

(1) Kanatzidis, M. G. *Acc. Chem. Res.* **2005**, *38* (4), 361–370.
(2) Bachhuber, F.; Krach, A.; Furtner, A.; Söhnel, T.; Peter, P.; Rothballer, J.; Weihrich, R. *J. Solid State Chem.* **2015**, *226*, 29–35.
(3) Bachhuber, F.; Rothballer, J.; Söhnel, T.; Weihrich, R. *Comput. Mater. Sci.* **2014**, *89*, 114–121.
(4) Fleurial, J.-P.; Caillat, T.; Borshchevsky, A. *XVI ICT "97. Proc. ICT"97. 16th Int. Conf. Thermoelectr. (Cat. No.97TH8291)* **1997**, 1–11.
(5) Uher, C. In *Chemistry, physics, and materials science of thermoelectric materials; beyond bismuth telluride*; 2003; pp 121–146.
(6) Zabel, M.; Wandinger, S.; Range, K. *Zeitschrift für Naturforsch. B* **1979**, *34b*, 238–241.
(7) Jackeli, G.; Khaliullin, G. *Phys. Rev. Lett.* **2009**, *102* (1), 017205.
(8) Shitade, A.; Katsura, H.; Kuneš, J.; Qi, X.-L.; Zhang, S.-C.; Nagaosa, N. *Phys. Rev. Lett.* **2009**, *102* (25), 256403.
(9) Gretarsson, H.; Clancy, J. P.; Liu, X.; Hill, J. P.; Bozin, E.; Singh, Y.; Manni, S.; Gegenwart, P.; Kim, J.; Said, A. H.; Casa, D.; Gog, T.; Upton, M. H.; Kim, H.-S.; Yu, J.; Katukuri, V. M.; Hozoi, L.; van den Brink, J.; Kim, Y.-J. *Phys. Rev. Lett.* **2013**, *110* (7), 076402.
(10) Chaloupka, J.; Jackeli, G.; Khaliullin, G. *Phys. Rev. Lett.* **2010**, *105* (2), 027204.
(11) Wallace, D. C.; Brown, C. M.; McQueen, T. M. *J. Solid State Chem.* **2015**, *224*, 28–35.
(12) Katukuri, V. M.; Nishimoto, S.; Yushankhai, V.; Stoyanova, A.; Kandpal, H.; Choi, S.; Coldea, R.; Rousochatzakis, I.; Hozoi, L.; Brink, J. van den. *New J. Phys.* **2014**, *16* (1), 013056.
(13) Lei, H.; Yin, W.-G.; Zhong, Z.; Hosono, H. *Phys. Rev. B* **2014**, *89* (2), 020409.
(14) Rau, J. G.; Lee, E. K.-H.; Kee, H.-Y. *Phys. Rev. Lett.* **2014**, *112* (7), 077204.
(15) Andrade, E. C.; Vojta, M. *Phys. Rev. B* **2014**, *90* (20), 205112.
(16) Kimchi, I.; Analytis, J. G.; Vishwanath, A. *Phys. Rev. B* **2014**, *90* (20), 205126.
(17) Munson, R. A. *Inorg. Chem.* **1968**, *7* (2), 389–390.
(18) Jobic, S.; Deniard, P.; Brec, R.; Rouxel, J.; Drew, M. G. B.; David, W. I. F. *J. Solid State Chem.* **1990**, *89* (2), 315–327.
(19) Jobic, S.; Deniard, P.; Brec, R.; Rouxel, J.; Jouanneaux, A.; Fitch, A. N. *Zeitschrift fuer Anorg. und Allg. Chemie* **1991**, *598/599* (1), 199–215.
(20) Jobic, S.; Brec, R.; Chateau, C.; Haines, J.; Léger, J. M.; Koo, H. J.; Whangbo, M. H. *Inorg. Chem.* **2000**, *39* (19), 4370–4373.
(21) Li, L.; Qi, T. F.; Lin, L. S.; Wu, X. X.; Zhang, X. T.; Butrouna, K.; Cao, V. S.; Zhang, Y. H.; Hu, J.; Yuan, S. J.; Schlottmann, P.; De Long, L. E.; Cao, G. *Phys. Rev. B* **2013**, *87* (17), 174510.
(22) Yang, J. J.; Choi, Y. J.; Oh, Y. S.; Hogan, A.; Horibe, Y.; Kim, K.; Min, B. I.; Cheong, S.-W. *Phys. Rev. Lett.* **2012**, *108* (11), 116402.
(23) Zhou, S. Y.; Li, X. L.; Pan, B. Y.; Qiu, X.; Pan, J.; Hong, X. C.; Zhang, Z.; Fang, A. F.; Wang, N. L.; Li, S. Y. *EPL* **2013**, *104* (2), 27010.
(24) Fang, A. F.; Xu, G.; Dong, T.; Zheng, P.; Wang, N. L. *Sci. Rep.* **2013**, *3*, 1153.
(25) Ootsuki, D.; Pyon, S.; Kudo, K.; Nohara, M.; Horio, M.; Yoshida, T.; Fujimori, A.; Arita, M.; Anzai, H.; Namatame, H.; Taniguchi, M.; Saini, Naurang, L.; Mizokawa, T. *J. Phys. Soc. Japan* **2013**, *82*, 093704.
(26) Momma, K.; Izumi, F. *J. Appl. Crystallogr.* **2008**, *41* (3), 653–658.
(27) The ELK FP-LAPW Code, (available at http://elk.sourceforge.net).
(28) Fu, L.; Kane, C. L. *Phys. Rev. B - Condens. Matter Mater. Phys.* **2007**, *76* (4), 1–17.
(29) Mostofi, A. A.; Yates, J. R.; Lee, Y.-S.; Souza, I.; Vanderbilt, D.; Marzari, N. *Comput. Phys. Commun.* **2008**, *178* (9), 685–699.
(30) Kresse, G.; Furthmüller, J. *Comput. Mater. Sci.* **1996**, *6* (1), 15–50.
(31) Kresse, G. *Phys. Rev. B* **1996**, *54* (16), 11169–11186.
(32) Kresse, G.; Hafner, J. *Phys. Rev. B* **1993**, *47* (1), 558–561.
(33) Gresch, D., Soluyanov, A. A., Vanderbilt, D., Autes, G., Yazyev, O., Ceresoli, D., Troyer, M. *Prep.*
(34) Soluyanov, A. A.; Vanderbilt, D. *Phys. Rev. B* **2011**, *83* (23), 235401.
(35) Brostigen, G.; Kjekshus, A.; Astrup, E. E.; Nordal, V.; Lindberg, A. a.; Craig, J. C. *Acta Chem. Scand.* **1969**, *23*, 2186–2188.
(36) Bayliss, P. *Am. Mineral.* **1982**, *67*, 1048–1057.
(37) Ramsdell, L. S. *J. Mineral. Soc. Am.* **1925**, *10* (9), 281–304.
(38) Hamilton, W. C. *Acta Cryst.* **1965**, *18*, 502–510.
(39) Giacovazzo, C.; Monaco, H. L.; Artioli, G.; Viterbo, D.; Milanesio, M.; Ferraris, G.; Gilli, G.; Gilli, P.; Zanotti, G.; Catti, M. *Fundamentals of Crystallography Third Edition*; 2011.

(40) Spek, A. L. *Acta Crystallogr. Sect. D Biol. Crystallogr.* **2009**, *65* (2), 148–155.
(41) Baur, W. H. *Acta Crystallogr. Sect. B Struct. Crystallogr. Cryst. Chem.* **1976**, *32* (7), 2200–2204.
(42) Ramirez, A.; Kowach, G. *Phys. Rev. Lett.* **1998**, 4903–4906.
(43) Tari, A. *The Specific Heat of Matter at Low Temperatures*; Imperial College Press: London, 2003.
(44) Fritzsche, H. *Phys. Rev.* **1955**, *99* (2), 406–419.
(45) Fritzsche, H.; Lark-Horovitz, K. *Phys. Rev.* **1954**, *20*, 834–844.
(46) Li, S. S. *Natl. Bur. Stand. Spec. Publ.* **1979**, *400-47*, 1–42.
(47) Ashcroft, N. W.; Mermin, N. D. *Solid State Physics*; 1976.
(48) Mott, N. F. *Rev. Mod. Phys.* **1968**, *40* (4), 677–683.
(49) Cottingham, P.; Miller, D. C.; Sheckelton, J. P.; Neilson, J. R.; Feygenson, M.; Huq, A.; McQueen, T. M. *J. Mater. Chem. C* **2014**, *2* (17), 3238.
(50) Pippard, A. B. *Magnetoreistance in Metals*; Cambridge University Press: Cambridge, 1989.
(51) Snyder, G. J.; Toberer, E. S. *Nat. Mater.* **2008**, *7* (2), 105–114.
(52) Tritt, T. M. *Thermal Conductivity: Theory, Properties, and Applications*; Kluwer Academic/Plenum Publishers: New York, 2004.

# Anion-Anion Bonding and Topology in Ternary Iridium Seleno-Stannides

Benjamin A. Trump,*[,a,b] Jake A. Tutmaher,[b,c] Tyrel M. McQueen[a,b,c,d]

[a]Department of Chemistry, Johns Hopkins University, Baltimore, Maryland 21218, United States

[b]Institute for Quantum Matter, Johns Hopkins University, Baltimore, Maryland 21218, United States

[c]Department of Physics and Astronomy, Johns Hopkins University, Baltimore, Maryland 21218, United States

[d]Department of Material Science, Johns Hopkins University, Baltimore, Maryland 21218, United States

**Supporting Methods.** Magnetization measurements were run on a Quantum Design Physical Property Measurement system using an applied field of $\mu_0 H = 1$ T. Phases on the ternary diagram were explored by heating appropriate stoichiometric ratios according to the main text. The final, reported crystallographic information was obtained using TOPAS (Bruker AXS).

**Supporting Results and Discussion.** Magnetization measurements (Figure S1) on $Ir_2SnSe_5$, $IrSn_{0.45}Se_{1.55}$, and $Ir_2Sn_3Se_3$ demonstrate diamagnetic behavior, as all three have a small, negative, temperature-independent response. This defends the assignment of low-spin $5d^6$ $Ir^{3+}$ for electron counting. All three compounds also have a Curie-tail, due to some amount of defects as well.

Figure S2 shows the known Ir-Sn-Se phases in a ternary diagram. $IrSn_{0.45}Se_{1.55}$ exists within the solid solution of $Ir_2Sn_{1-\delta}Se_{3-\delta}$, however it competes with the more thermodynamically stable $Ir_2Sn_3Se_3$ and $IrSe_2$. The proposed tie lines in Figure S2 demonstrate that $IrSn_{0.45}Se_{1.55}$ is accessible because it lies outside the phase field of the more stable products.

Tables S1-S3 present the crystallographic information for $IrSn_{0.45}Se_{1.55}$, $Ir_2Sn_3Se_3$, and $Ir_2SnSe_5$ respectively, for the crystals structures described in the text. Occupancies were fixed due to refining within 1% of nominal values and errors given are statistical uncertainties for all three compounds.

Representative calculations $Ir_2SnSe_5$ and $IrSn_{0.45}Se_{1.55}$ are shown in Figures S3 and S4 respectively, using the methods described in the text, with a 10 x 10 x 6 and a 4 x 4 x 4 k-mesh respectively. $Ir_2SnSe_5$ is seen to be an indirect gap semiconductor, while $IrSn_{0.45}Se_{1.55}$ is seen to be a semimetal. $Ir_2SnSe_5$ converged within an absolute energy difference of $10^{-3}$ Ha whereas all other calculations converged with an absolute energy difference of ~$10^{-4}$ Ha. The $IrSn_{0.45}Se_{1.55}$ calculation used an ordered, more stoichiometric, lower symmetry $P2_1$ model, which is not fully representative of the higher symmetry, off-stoichiometric, disordered structure.

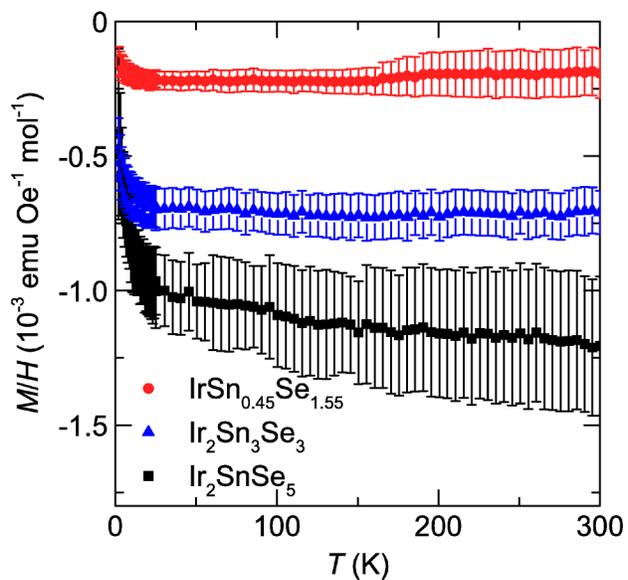

**Figure S1.** Magnetization versus temperature for $IrSn_{0.45}Se_{1.55}$ (red circles), $Ir_2Sn_3Se_3$ (blue triangles), and $Ir_2SnSe_5$ (black squares).

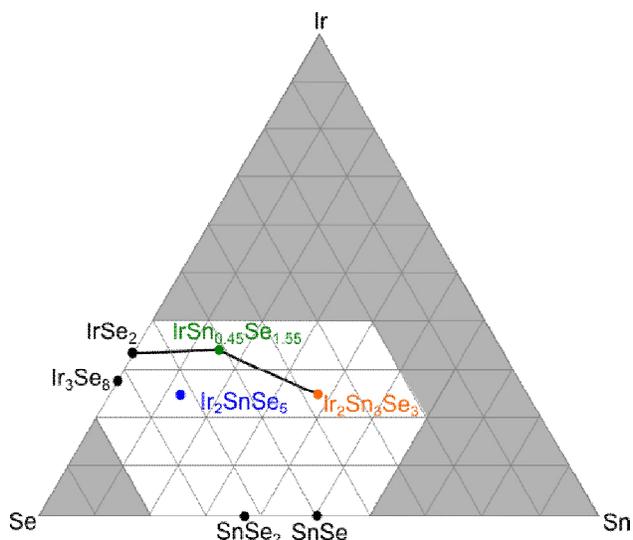

**Figure S2.** Ternary diagram for known Ir-Sn-Se compounds with a few proposed tie lines. Shaded areas are unexplored.

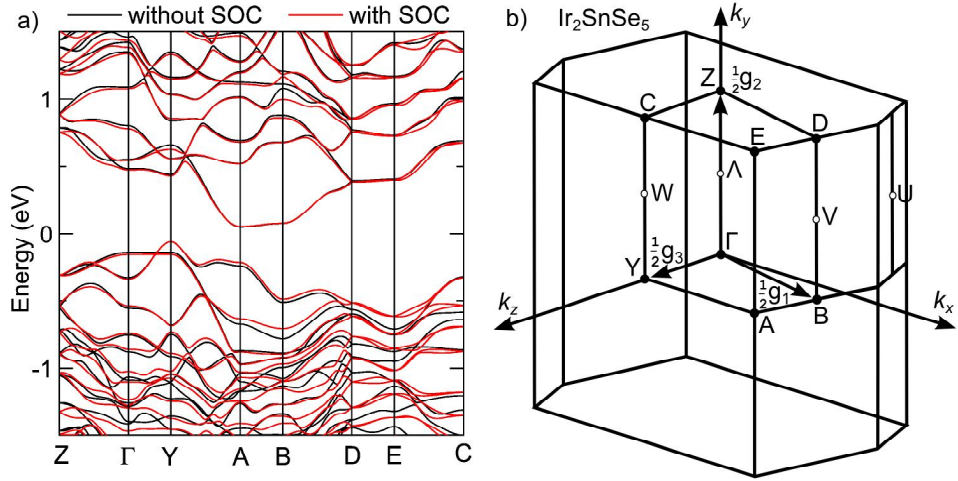

**Figure S3.** a) Band structure for $Ir_2SnSe_5$ without (black) and with (red) spin-orbit coupling (SOC). b) The Brillioun zone for $P2_1/m$ $Ir_2SnSe_5$ with special points and reciprocal lattice vectors shown.

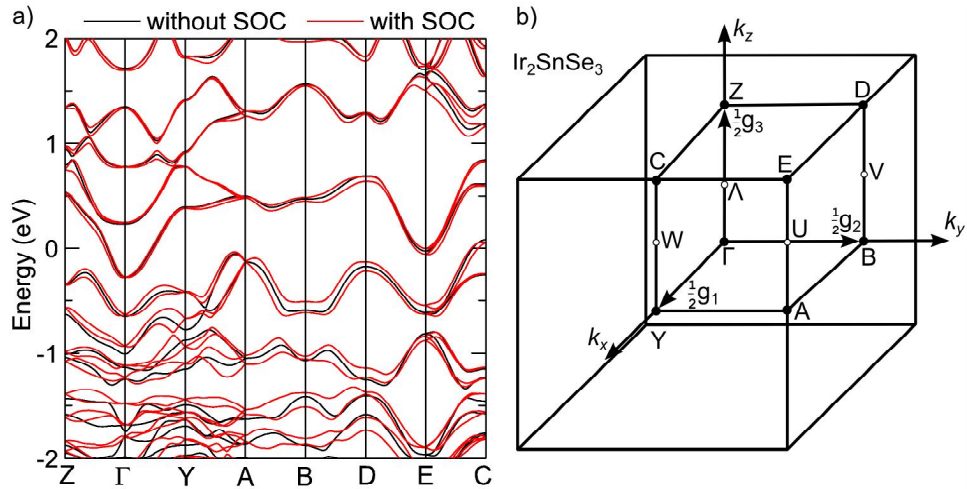

**Figure S4.** a) Representative band structure for $IrSn_{0.45}Se_{1.55}$ without (black) and with (red) spin-orbit coupling (SOC) using an ordered $P2_1$ model. b) The Brillioun zone for an ordered $P2_1$ $IrSn_{0.5}Se_{1.5}$ model.

S4

**Table S1**. Crystallographic parameters for IrSn$_{0.45}$Se$_{1.55}$ using $Pa3$ (205) obtained from Rietveld refinements to laboratory powder diffraction data at room temperature. Atomic positions are restricted by symmetry as Ir: 4$a$ (0, 0, 0) and Sn/Se: 8$c$ ($x$, $x$, $x$). Occupancies were fixed at nominal values and errors reported are from statistical uncertainties.

|       |                     |             |
|-------|---------------------|-------------|
|       | $\lambda$ (Å)       | 1.5418      |
|       | $a=b=c$ (Å)         | 6.074419(4) |
|       | $V$ (Å$^3$)         | 224.1374(5) |
| Ir    | $U_{iso}$ (Å$^2$)   | 0.00793(5)  |
|       | $x$                 | 0.37629(4)  |
| Sn/Se | occ                 | 0.225/0.775 |
|       | $U_{iso}$ (Å$^2$)   | 0.00488(8)  |
|       | $R_{wp}$            | 5.534       |
|       | $R_p$               | 4.157       |
|       | $R_F^2$             | 2.964       |
|       | $\chi^2$            | 1.867       |

$$R_{wp} = \sqrt{\frac{\sum |Y_{o,m} - Y_{c,m}|}{\sum Y_{o,m}}}$$

$$R_p = \sqrt{\frac{\sum w_m (Y_{o,m} - Y_{c,m})^2}{\sum w_m Y_{o,m}^2}}$$

$$R_{exp} = \sqrt{\frac{\sum M - P}{\sum w_m Y_{o,m}^2}}$$

$$GoF = \chi^2 = \sqrt{\frac{\sum w_m (Y_{o,m} - Y_{c,m})^2}{M - P}}$$

**Table S2**. Crystallographic parameters for Ir$_2$Sn$_3$Se$_3$ using rhombohedral $R3$ (148) obtained from Rietveld refinement of synchrotron powder diffraction data at room temperature. Atoms are restricted by symmetry as 2$c$ ($x$, $x$, $x$) and 6$f$ ($x$, $y$, $z$). Atomic displacement parameters ($U_{iso}$) for Sn and Se were constrained with each other and occupancies were fixed at nominal values and errors reported are from statistical uncertainties.

|                             |              |           |           |
|-----------------------------|--------------|-----------|-----------|
| $\lambda$ (Å)               | 0.41385      | $R_{wp}$  | 8.646     |
| $a = b = c$ (Å)             | 8.955798(6)  | $R_p$     | 7.266     |
| $\alpha = \beta = \gamma$ (°) | 89.92625(5) | $R_{exp}$ | 8.052     |
| $V$ (Å$^3$)                 | 718.3098(14) | GoF       | 1.074     |

| Atom | Wyck. Pos. | $x$         | $y$          | $z$         | $U_{iso}$ (Å$^2$) |
|------|------------|-------------|--------------|-------------|-------------------|
| Ir1  | 2$c$       | 0.24568(9)  | 0.24568(9)   | 0.24568(9)  | 0.0048(2)         |
| Ir2  | 6$f$       | 0.74433(9)  | 0.24428(10)  | 0.74537(9)  | 0.00522(9)        |
| Sn1  | 6$f$       | 0.16794(13) | 0.5001(2)    | 0.34972(14) | 0.007338(10)      |
| Sn2  | 6$f$       | 0.33250(13) | 0.99841(12)  | 0.84959(14) | 0.007338(10)      |
| Se1  | 6$f$       | 0.8527(2)   | 0.6531(2)    | 0.9990(2)   | 0.006421(14)      |
| Se2  | 6$f$       | 0.6499(2)   | 0.1518(2)    | 0.5000(2)   | 0.006421(14)      |

**Table S3**. Crystallographic parameters for $Ir_2SnSe_5$ using $P2_1/m$ (11) obtained from Rietveld refinement of synchrotron powder diffraction data at room temperature. Atoms are restricted by symmetry as 2e (x, 1/4, z) and 4f (x y z). Atomic displacement parameters ($U_{iso}$) for Sn and Se were constrained with each other. Occupancies were fixed at nominal values and errors reported are from statistical uncertainties.

| λ (Å) | 0.41388 | α = γ (°) | 90 | $R_{wp}$ | 12.078 |
|---|---|---|---|---|---|
| a (Å) | 7.65768(5) | β (°) | 102.0831(6) | $R_p$ | 9.791 |
| b (Å) | 7.51027(5) | V (Å³) | 702.815(8) | $R_{exp}$ | 5.200 |
| c (Å) | 12.49737(8) | | | GoF | 2.323 |

| atom | Wyck. Pos. | x | y | z | $U_{iso}$ (Å²) |
|---|---|---|---|---|---|
| Ir1 | 4f | 0.7439(20) | -0.0015(5) | 0.15063(10) | 0.00567(14) |
| Ir2 | 2e | 0.2475(3) | 1/4 | 0.1395(2) | 0.00567(14) |
| Ir3 | 2e | 0.7474(3) | 1/4 | 0.8364(2) | 0.00567(14) |
| Se1 | 4f | 0.0447(3) | -0.0038(11) | 0.1091(2) | 0.0106(2) |
| Se2 | 4f | 0.5412(4) | -0.0041(10) | 0.7821(2) | 0.0106(2) |
| Se3 | 2e | 0.3306(13) | 1/4 | 0.9723(5) | 0.0106(2) |
| Se4 | 2e | 0.6677(13) | 1/4 | 0.0222(5) | 0.0106(2) |
| Se5 | 2e | 0.8161(10) | 1/4 | 0.3004(4) | 0.0106(2) |
| Se6 | 2e | 0.1833(10) | 1/4 | 0.7165(5) | 0.0106(2) |
| Se7 | 4f | 0.7498(4) | -0.0023(10) | 0.5202(2) | 0.0106(2) |
| Sn1 | 2e | 0.8352(8) | 1/4 | 0.6513(3) | 0.0089(4) |
| Sn2 | 2e | 0.1651(8) | 1/4 | 0.3360(3) | 0.0089(4) |

**Table S4**. Mobilities extracted from Hall ($\mu_H$) and Resistivity ($\mu$) data.

| T (K) | $\mu_H$ (cm²V⁻¹s⁻¹) | $\mu$ (cm²V⁻¹s⁻¹) |
|---|---|---|
| 4 | 0.84(8) | 0.10(1) |
| 15 | 0.95(9) | 0.078(7) |
| 40 | 0.88(8) | 0.030(3) |
| 150 | 3.6(3) | 0.40(4) |
| 300 | 11(1) | 1.9(2) |